\documentclass{ws-ijmpcs}

\def\mnras{MNRAS}
\def\apjl{ApJL}
\def\apj{ApJ}

\def\simless{\mathbin{\lower 3pt\hbox
{$\rlap{\raise 5pt\hbox{$\char'074$}}\mathchar"7218$}}}   
\def\simmore{\mathbin{\lower 3pt\hbox
{$\rlap{\raise 5pt\hbox{$\char'076$}}\mathchar"7218$}}}   

\begin{document}

\markboth{Mimica \& Aloy}
{Efficiency of internal shocks in magnetized relativistic jets}

%
\catchline{}{}{}{}{}
%

\title{EFFICIENCY OF INTERNAL SHOCKS IN MAGNETIZED RELATIVISTIC JETS}

\author{PETAR MIMICA}

\address{Departament d'Astronomia i Astrofísica, Universitat de
  Valencia\\ C/ Dr. Moliner, 50, 
  E-46100 Burjassot (Valencia), Spain\\
Petar.Mimica@uv.es}

\author{MIGUEL ANGEL ALOY}

\address{Departament d'Astronomia i Astrofísica, Universitat de
  Valencia\\ C/ Dr. Moliner, 50, 
  E-46100 Burjassot (Valencia), Spain\\
Miguel.A.Aloy@uv.es}

\maketitle

\begin{history}
\received{Day Month Year}
\revised{Day Month Year}
\end{history}

\begin{abstract}
  We study the dynamic and radiative efficiency of conversion of
  kinetic-to-thermal/magnetic energy by internal shocks in
  relativistic magnetized outflows. A parameter study of a large
  number of collisions of cylindrical shells is performed. We explore
  how, while keeping the total flow luminosity constant, the variable
  fluid magnetization influences the efficiency and find that the
  interaction of shells in a mildly magnetized jet yields higher
  dynamic, but lower radiative efficiency than in a non-magnetized
  flow. A multi-wavelength radiative signature of different shell
  magnetizations is computed assuming that relativistic particles are
  accelerated at internal shocks.
\end{abstract}

\keywords{Magnetohydrodynamics (MHD); Shock waves; radiation
  processes: non-thermal; gamma-ray bursts; relativistic jets: blazars}

\section{Introduction}	

The radiation observed in the relativistic outflows of blazars and
gamma-ray bursts (GRBs) shows a high degree of variability. Although
the radiation energy and time scales are different for both classes of
objects, the underlying physics is probably very similar. The internal
shock scenario\cite{Rees:1994ca} has been used to explain the
variability of blazars\cite{Spada:2001do,Mimica:2004ay} and
GRBs:\cite{Kobayashi:1997vf,Daigne:1998wq,Bosnjak:2009dv}
inhomogeneities in the outflow cause different parts (shells) to
collide and produce the internal shock waves. The dissipation
originated at such shocks has been studied by means of relativistic
(magneto)hydrodynamic
simulations\cite{Kino:2004in,Mimica:2004ay,Mimica:2005aa,Mimica:2007aa}
in non-/weakly magnetized outflows, and has shown the complexity of
shell interactions, not taken fully into account by simpler analytic
models.\cite{Kobayashi:1997vf,Daigne:1998wq,Spada:2001do,Bosnjak:2009dv}

These studies did not address the highly magnetized regime of shell
interactions, particularly, they did not consider whether it is only
kinetic or also the magnetic energy which can be used to power the
emission. In a previous work\cite{Mimica:2010aa} we study the
efficiency of conversion of the kinetic to thermal and magnetic energy
(we call it \emph{dynamic} efficiency). The main results of that study
show that 1) the dynamic efficiency is highest if the shells are
moderately magnetized (magnetization $\sigma \approx 0.1$), 2) if the relative
difference in magnetization between colliding shells is $\approx 5 -
100$, the efficiency of a single collision can be as high as $20$\% and
3) multiple shell collisions can raise the efficiency to even higher
values (close to $100$\%).

In this work we discuss some preliminary results of the upcoming
paper\cite{Mimica:2011aa} on the \emph{radiative} efficiency of the
shell collisions. More precisely, we discuss the definitions of the
efficiency and the radiative signature of strongly magnetized shell
collisions.

\section{Dynamic and radiative efficiency}
\label{sec:eff}

In our previous work\cite{Mimica:2010aa}, we define the dynamic
thermal and magnetic efficiency ($\varepsilon_T$ and $\varepsilon_M$,
respectively) as a fraction of the total initial energy in the shells
which is converted into an thermal and magnetic energy in the merged
shell, respectively. We also defined the total dynamic efficiency as
$\varepsilon:= \varepsilon_T + \varepsilon_M$.

The radiative efficiency $\varepsilon_R$ can be defined as a fraction
of the total initial energy in the shells converted into
radiation. Assuming that only a fraction $\varepsilon_e$ of the
thermal energy dissipated at the shock can be transferred to
non-thermal particles and later radiated, we can write
\begin{equation}
  \varepsilon_R = f_R \varepsilon_e \varepsilon\, 
\end{equation}
where $f_R:= \varepsilon_T / (\varepsilon_T + \varepsilon_M)$. The
reason that we only take the thermal energy as being available for
radiation is that we do not consider any magnetic dissipation
processes (ideal limit).
\begin{figure}
\centering 
\includegraphics[scale=0.3]{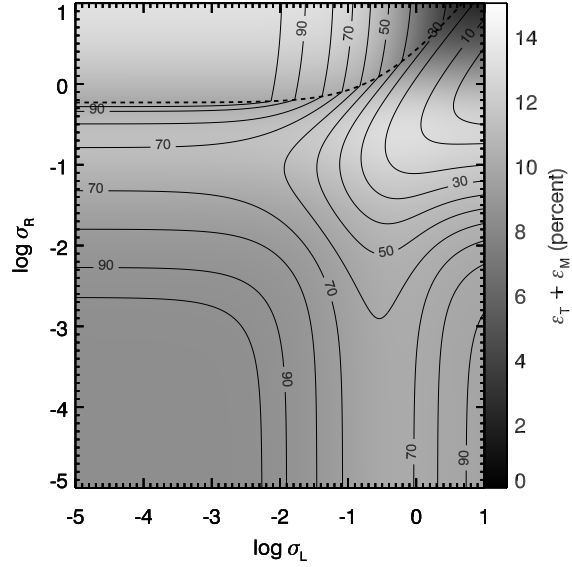}
\caption[]{Contours of $f_R$ (solid lines) plotted over the dynamic
  efficiency distribution (grey shades) as a function of left (faster)
  and right (slower) shell magnetizations. The contours show $f_R$ in
  per cent with the values $1$, $5$, $10$, $20$, $30$, $40$, $50$,
  $60$, $70$, $80$, $90$, $95$ and $100$. A value
  $\varepsilon_e=1$ is considered. }
\label{fig:eff}
\end{figure}
Figure~\ref{fig:eff} shows the contours of $f_R$ and the filled
contours of $\varepsilon$ as a function of the shell magnetization. We
denote the faster and slower shells by the subscripts $L$ and $R$,
respectively. The shell Lorentz factors are $\Gamma_L = 20$ and
$\Gamma_R = 10$, which corresponds to a typical blazar jet. It can be
seen that, although the dynamic efficiency has its maximum in the
region $(\sigma_L \simeq 1,\ \sigma_R \simeq 0.1)$, there, the
radiative efficiency is lower than for the weakly magnetized shells
($f_R\leq 0.3$ there, as opposed to $f_R\geq 0.9$ when
$\sigma_{L,R}\leq 0.01$). The plot has been obtained using a grid of
$10^6$ different values of $(\sigma_L, \sigma_R)$ pairs and then
solving Riemann problems using method described in our previous
work\cite{Mimica:2010aa}.

\section{Radiative signature}
\label{sec:rad}

\begin{figure}
\centering 
\includegraphics[scale=0.3]{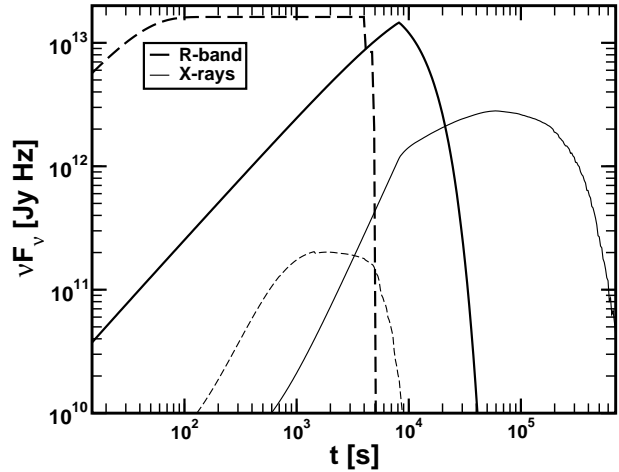}
\caption[]{Optical (thick lines) and X-ray (thin lines) light
  curves for the models $(\sigma_L = 10^{-6}, \sigma_R = 10^{-6})$
  (full lines) and $(\sigma_L = 1, \sigma_R = 0.1)$ (dashed lines). }
\label{fig:rad}
\end{figure}

Figure~\ref{fig:rad} shows the light curves for two models, $(\sigma_L
= 10^{-6}, \sigma_R = 10^{-6})$ (full lines) and $(\sigma_L = 1,
\sigma_R = 0.1)$ (dashed lines). Shown are the optical (thick lines)
and X-ray (thin lines) light curves.
The method for the computation of emission will be described in our
upcoming article\cite{Mimica:2011aa}, and is very similar to the ones
already used to study multiwavelength light curves from non-magnetized
shell collisions\cite{Bottcher:2010gn,Joshi:2011bp}. We take into
account the synchrotron and the synchrotron self-Compton (SSC)
processes. The optical radiation is dominated by the synchrotron
process, while the X-ray radiation is dominated by the SSC. Once the
shocks cross the shells the optical emission peaks and drops
afterward, while the SSC emission peaks at later times because of the
delay caused by the travel time of the seed synchrotron photons, which
are scattered to X-ray frequencies.

The difference between the non-magnetized (full lines) and the
magnetized model (dashed lines) is both in the intensity of the emitted
radiation, as well as in the shape of the light curve. The magnetized
model optical light curve is almost constant until the sharp drop-off,
corresponding to the point when the shocks cross the shells. Due to
the high magnetic field the electrons are fast cooling and the
radiation drops very soon after the shocks stop accelerating fresh
particles. In the non-magnetized case the magnetic field is low enough
that the electrons are slow cooling and the light curve peaks when the
shocks cross the shell, and then decays gently as the electrons cool.
%
 The SSC emission from the magnetized model is
much lower than the one from the non-magnetized one due to the lower
density of the scattering electrons (both models have the same total
energy, but in the magnetized case a substantial part of the energy is
carried by the magnetic field, while in the non-magnetized case all
the energy is in the baryons).

\section{Conclusions}
\label{sec:conclusions}

We studied the radiative efficiency and the radiative signature of the
collisions of arbitrarily magnetized shells in the variable outflow
of blazars and GRBs. Our results are preliminary, but they show that,
while the dynamic efficiency in magnetized outflows can be higher, the
radiative efficiency (at least in the internal shocks model) is
probably much lower. The multi-wavelength light curve also shows
substantial differences between the two cases, notably in the shape of
the optical light curve.

\section*{Acknowledgments}
PM and MA acknowledge the support from the European Research Council
(grant CAMAP-259276), and the partial support of grants
AYA2007-67626-C03-01, CSD2007-00050, and PROMETEO-2009-103. The
authors thankfully acknowledge the computer resources, technical
expertise and assistance provided by the Barcelona Supercomputing
Center - Centro Nacional de Supercomputaci\'on.


\begin{thebibliography}{00}
\bibitem{Rees:1994ca} Rees M.~J., Meszaros P., {\it \apjl} {\bf
    430}, L93 (1994).
\bibitem{Spada:2001do} Spada M., Ghisellini G., Lazzati D., Celotti A., {\it \mnras} {\bf 325}, 1559 (2001).
\bibitem{Mimica:2004ay} Mimica P., Aloy M.~A., M{\"u}ller E.,
  Brinkmann W., {\it A{\&}A} {\bf 418}, 947 (2004)
\bibitem{Kobayashi:1997vf} Kobayashi S., Piran T., Sari R., {\it
    \apj} {\bf 490}, 92 (1997)
\bibitem{Daigne:1998wq} Daigne F., Mochkovitch R., {\it \mnras}
  {\bf 296}, 275 (1998)
\bibitem{Bosnjak:2009dv} Bo{\v s}njak {\v Z} , Daigne F., Dubus G.,
  {\it A{\&}A} {\bf 498}, 677 (2009)
\bibitem{Kino:2004in} Kino M., Mizuta A., Yamada S., {\it \apj}
  {\bf 611}, 1021 (2004)
\bibitem{Mimica:2005aa} Mimica P., Aloy M.~A., M{\"u}ller E.,
  Brinkmann W., {\it A{\&}A} {\bf 441}, 103 (2005)
\bibitem{Mimica:2007aa} Mimica P., Aloy M.~A., M{\"u}ller E.,
  {\it A{\&}A} {\bf 466} 93 (2007)
\bibitem{Mimica:2010aa} Mimica P., Aloy M.~A., {\it \mnras} {\bf
    401}, 525 (2010)
\bibitem{Mimica:2011aa} Mimica P., Aloy M.~A., in preparation (2011)
\bibitem{Bottcher:2010gn} B{\"o}ttcher M., Dermer C., {\it \apj}
  {\bf 711}, 445 (2010)
\bibitem{Joshi:2011bp} Joshi M., B{\"o}ttcher M., {\it \apj} {\bf
    727}, 21 (2011)
\end{thebibliography}
\end{document}